\documentclass[11pt, twocolumn, twoside]{IEEEtran}
\usepackage{cite}
\usepackage[cmex10]{amsmath}
\usepackage{fixltx2e}
\usepackage{datetime}
\usepackage{graphicx}
\usepackage{stfloats}
\usepackage{color}
\usepackage[T1]{fontenc}
\usepackage[utf8]{inputenc}
\usepackage{authblk}
\usepackage{amssymb}
\usepackage{algorithm}
\usepackage{algorithmic}
\usepackage{pstricks,epsfig}
\usepackage{rotating}
\usepackage{subfig}
\usepackage{tikz}
\usepackage{enumerate}
\usepackage{tabularx}
\usepackage{verbatim}
\usepackage{pgf}
\usepackage{url}
\usepackage[margin=1.75cm]{geometry}

\makeatletter

\newcommand{\Rmnum}[1]{\expandafter\@slowromancap\romannumeral #1@}
\makeatother

\makeatletter
\def\BState{\State\hskip-\ALG@thistlm}
\makeatother

\begin{document}
%
\title{A Bayesian Residual Transform for Signal Processing}
%
%
%

\author{Alexander~Wong, \textit{IEEE Member} and Xiao~Yu~Wang
\thanks{The authors are with the Vision and Image Processing Lab, Department
of Systems Design Engineering, University of Waterloo, 200 University Ave. West, Waterloo, Ontario, Canada, N2L 3G1. Tel.: +1 519 888 4567 x35342. Fax: +1 519 746 4791. E-mail: \{a28wong, x18wang\}@uwaterloo.ca.}}

%
%

\markboth{IEEE Access journal}%
{A Bayesian Residual Transform for Signal Processing}
%



\maketitle

\begin{abstract}
Multi-scale decomposition has been an invaluable tool for the processing of physiological signals.  Much focus in multi-scale decomposition for processing such signals have been based on scale-space theory and wavelet transforms.  In this study, we take a different perspective on multi-scale decomposition by investigating the feasibility of utilizing a Bayesian-based method for multi-scale signal decomposition called Bayesian Residual Transform (BRT) for the purpose of physiological signal processing.  In BRT, a signal is modeled as the summation of residual signals, each characterizing information from the signal at different scales.  A deep cascading framework is introduced as a realization of the BRT.  Signal-to-noise ratio (SNR) analysis using electrocardiography (ECG) signals was used to illustrate the feasibility of using the BRT for suppressing noise in physiological signals.  Results in this study show that it is feasible to utilize the BRT for processing physiological signals for tasks such as noise suppression.
\end{abstract}


%
\IEEEpeerreviewmaketitle

\section{Introduction}

Physiological signals are signals that are measured from sensors that are either placed on or implanted into the body.  Such physiological signals include those obtained using electromyography (EMG), electrocardiography (ECG), electroencephalography (EEG), photoplethysmography (PPG), and ballistocardiography (BCG).  The processing and interpretation of such signals is challenging due to a number of different factors.  For example, it is often difficult to obtain high-fidelity physiological signals due to noise, resulting in low signal-to-noise ratio (SNR).  Traditionally, signal averaging and linear filters such as band-reject and band-pass filters have been used to process such physiological signals to suppress noise; however, such approaches have also been shown to result in signal degradation~\cite{Christov,Mewette}.  As such, more advanced methods for handling such physiological signals are desired.

Multi-scale decomposition has become an invaluable tool for the processing of physiological signals.  In multi-scale decomposition, a signal is decomposed into a set of signals, each characterizing information about the original signal at a different scale.  A common signal processing task that multi-scale decomposition has shown to provide significant benefits is noise suppression, based on the notion that the information pertaining to the noise component would be largely characterized by certain scales that are separate from the scales characterizing the desired signal.  Much of literature in multi-scale decomposition for physiological signal processing has focused on scale-space theory~\cite{Melo08,Jager,Witkin83,Koenderink,Perona90,Gilboa08,scale1,scale2} and wavelet transforms~\cite{Phinyomark,Kestler,Hussain,Sobahi,Popescu,Chouakri,Agante,Donoho1,Donoho2}, with some investigations also conducted using methods such as empirical mode decomposition~\cite{Kopsinis,Weng,Velasco}.

In scale-space theory~\cite{Witkin83}, a signal $f(t)$ is decomposed into a single-parameter family of $n$ signals, denoted by $L$, with a progressive decrease in fine scale signal information between successive scales:

\begin{equation}
L = \{l_{j}(t) | 0 \leq j \leq n-1\},
\end{equation}

\noindent where $t$ denotes time, $j$ denotes scale, $l_{j}(t)$ is the signal at the ${j}^{\rm th}$ scale, and $l_0(t)=f(t)$.  By decomposing a signal into a set of signals with a progressive decrease in fine scale signal information between successive scales, one can then analyze signals at coarser scales without the influence of fine scale signal information such as that pertaining to noise, which is mainly characterized at the finer scales.  As such, one can utilize scale space theory to suppress noise in a signal by perform scale space decomposition on the signal and then treating one of signals at a coarser scale as the noise-suppressed signal.  However, there are several limitations to the use of scale space theory for physiological signal processing pertaining to noise suppression.  First, noise suppression using scale space theory requires the careful selection of which scale represents the noise-suppressed signal, which can be challenging.  Second, noise suppression using scale space theory does not facilitate for fine-grained noise suppression at the individual scales, which limits its overall flexibility in striking a balance between noise suppression and signal structural preservation.

In wavelet decomposition~\cite{Mallat89,Daubechies92}, a signal $f(t)$ is decomposed into a set of wavelet coefficients $c_{j,k}(t)$ obtained using a wavelet transform $W$:

\begin{equation}
c_{j,k}(t)=W_{\psi,f}(a,b)(t)
\end{equation}

\noindent where $\psi$ is the wavelet, $a=2^{-j}$ is the dyadic dilation, and $b=k2^{-j}$ is the dyadic position.  Wavelet transforms has a number of advantages for the purpose of physiological signal processing, particularly pertaining to noise suppression.  First, as signal information at different scales are better separated in the wavelet domain (i.e., signal information at one scale is not contained in another scale), this facilitates fine-grained noise suppression at the individual scales to strike a balance between noise suppression and signal structural preservation.  Second, scale selection when performing noise suppression using wavelet transforms is less critical than that for noise suppression using scale space theory, since all scales are considered in noise suppression using wavelet transforms as opposed to a single scale selection with scale space theory.  One limitation worth noting pertaining to signal processing using wavelet transforms, particularly pertaining to noise suppression, is that signals processed using wavelet transforms can exhibit oscillation artifacts related to wavelet basis functions used in the wavelet transform, which is particular noticeable when dealing with low SNR scenarios.  Therefore, given some of the limitations with both scale space theory and wavelet transforms when used for physiological signal processing, one is motivated to explore alternative approaches that can address these limitations.

Here, we take a different approach by exploring a Bayesian perspective to multi-scale signal decomposition.  In this perspective, a signal is viewed as an amalgamation of a number of signals, each characterizing unique signal information at a different scale with different statistical characteristics.  Taking such a perspective to the problem of multi-scale signal decomposition has a number of advantages.  First, like the wavelet transform, since signal information at one scale is not contained in another scale, it allows us to achieve the benefits of taking better advantage of fine-grained noise suppression at the individual scales to strike a balance between noise suppression and signal structural preservation.  Second, since signals are decomposed based on their statistical characteristics as opposed to a set of deterministic basis functions, signals processed using this approach would not exhibit the types of basis-related artifacts associated with the use of wavelet transforms.  Motivated by this, in this study, we investigate the feasibility of utilizing a new Bayesian-based method for multi-scale signal decomposition called Bayesian Residual Transform (BRT) for the purpose of physiological signal processing.

This paper is organized as follows.  First, the methodology behind the proposed Bayesian Residual Transform is described in Section~\ref{methods}.  The experimental setup for evaluating the feasibility of using the BRT for suppressing noise in physiological signals via signal-to-noise ratio (SNR) analysis using electrocardiography (ECG) signals is described in Section~\ref{setup}.  The experimental results and discussion is presented in Section~\ref{results}, and conclusions are drawn and future work discussed in Section~\ref{conclusions}.

\section{Bayesian Residual Transform}
\label{methods}

\begin{figure*}[tp]
	\centering
    \includegraphics[width=0.6\linewidth]{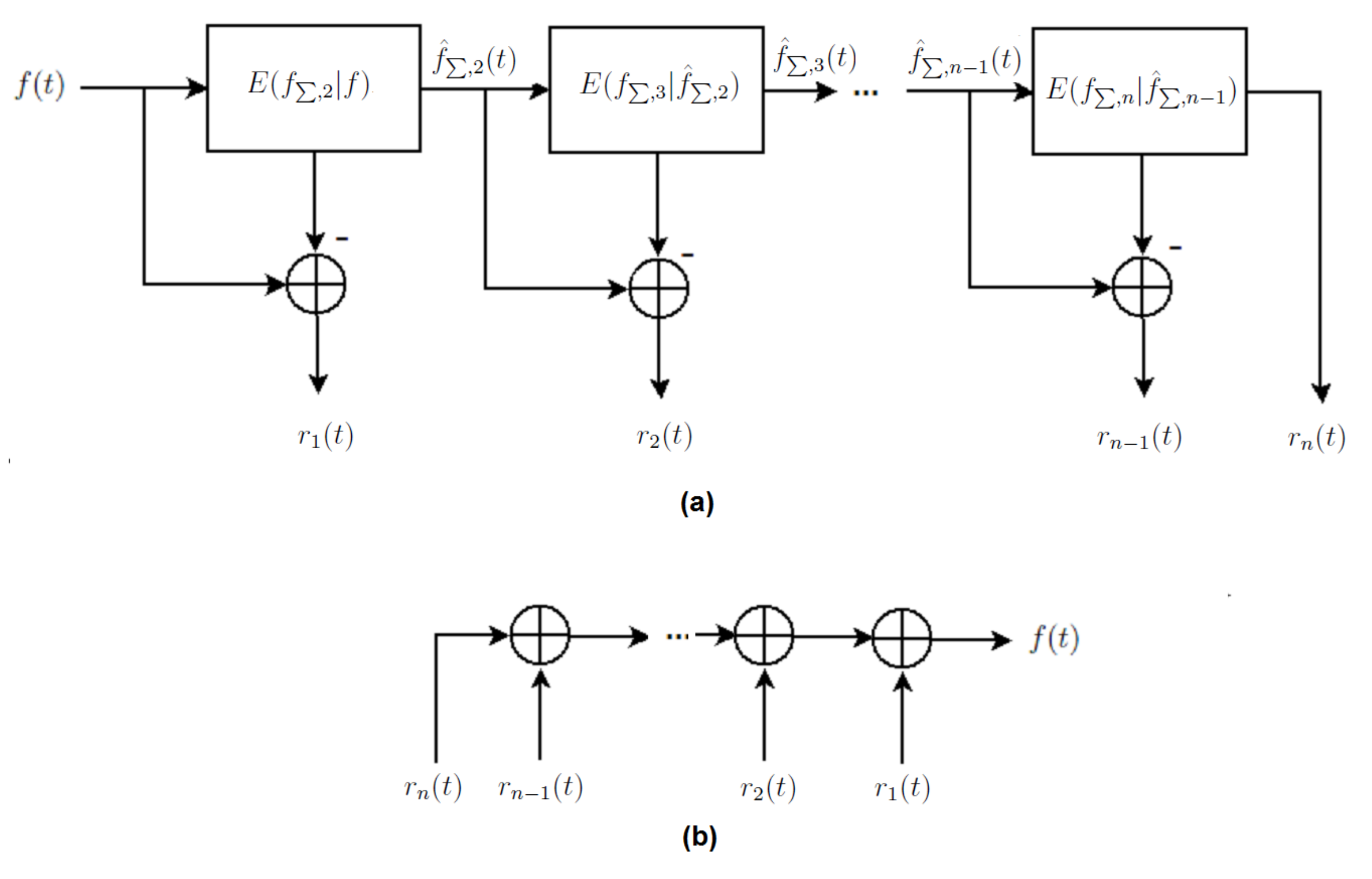}
	\caption{Bayesian Residual Transform framework. \textbf{(a)} forward BRT. \textbf{(b)} inverse BRT.}
	\label{fig1}
\end{figure*}

A full derivation of the proposed Bayesian Residual Transform (BRT) can be described as follows.  In the BRT, a signal $f(t)$ is modeled as the summation of $n$ residual signals, each characterizing signal information from the signal at increasingly coarse scales:

\begin{equation}
f(t) = \sum_{i=1}^{n}r_i(t) = f_{\sum,1}(t),
\label{sumofprocesses}
\end{equation}

\noindent  where $f_{\sum,j}(t)$ denote a signal representing the summation of all residual signals at scales $[j,n]$:
\begin{equation}
f_{\sum,j}(t) = \sum_{i=j}^{n}r_i(t),
\label{fsum}
\end{equation}

\noindent and $r_{i}=\{r_{i}(t) | t \in T\}$ is a residual signal characterizing the signal information at the $i^{\rm th}$ scale with different statistical characteristics.  The residual signals at the lower scales contain fine-grained signal characteristics of the signal, while the residual signals at the higher scales contain coarse-grained signal characteristics of the signal (e.g., $r_1(t)$ characterizes the finest-grained signal characteristics, while $r_n(t)$ characterizes the coarsest-grained signal characteristics).  As such, each residual signal contains unique information about the scale corresponding to a particular scale that the other residual signals do not contain.  Therefore, the goal of the BRT (denoted by the function $B$) is to decompose a signal $f(t)$ into the set of $n$ residual signals $r_1(t), r_2(t), \ldots, r_n(t)$:

  \begin{equation}
  \{r_1(t), r_2(t), \ldots, r_n(t)\}=B(f(t)).
  \end{equation}

  \noindent Determining the set of residual signals characterizing the signal information at the different scales and whose sum is equal to $f(t)$ (i.e., Eq.~\ref{sumofprocesses}) is a highly challenging problem, and as such with the BRT we wish to introduce a deep cascading framework to solve this problem in a more tractable manner, where a residual signal at a particular scale is computed based on computations performed at a previous scale.

  Let us first rewrite Eq.~\ref{sumofprocesses} as follows:

\begin{equation}
f_{\sum,1}(t) =  f_{\sum,2}(t) + r_1(t).
\label{process2}
\end{equation}

\noindent It can be observed from Eq.~\ref{process2} that the residual signal $r_1(t)$ can be treated as the residual between the summation of all residual signals at scales $[1,n]$ and the summation of all residual signals at scales $[2,n]$.  Hence, one can treat this as an inverse problem of estimating $f_{\sum,2}(t)$ given $f_{\sum,1}(t)$, with the analytical solution given by the conditional expectation $E(f_{\sum,2}(t)|f_{\sum,1}(t))$~\cite{Fieguth} (the quantification of the conditional expectation will be explained in more detail in a later section discussing the realization of the BRT via kernel regression).  Therefore, given ${\hat {f}}_{\sum,2}(t)=E(f_{\sum,2}(t)|f_{\sum,1}(t))$, one can substitute ${\hat {f}}_{\sum,2}(t)$ for ${f}_{\sum,2}(t)$ in Eq.~\ref{process2} and rearrange the terms to obtain $r_1(t)$ as:

\begin{equation}
r_1(t) = f_{\sum,1}(t) - {\hat {f}}_{\sum,2}(t).
\label{process5}
\end{equation}

\noindent Given ${\hat {f}}_{\sum,2}(t)$, which is computed to obtain $r_1(t)$, we can express the relationship between $r_2(t)$ and ${\hat {f}}_{\sum,2}(t)$ in a similar manner to Eq.~\ref{process2} as:

\begin{equation}
{\hat {f}}_{\sum,2}(t) =  f_{\sum,3}(t) + r_2(t),
\label{process5c}
\end{equation}

\noindent which can similarly be treated as an inverse problem of estimating $f_{\sum,3}(t)$ given ${\hat {f}}_{\sum,2}(t)$, with the analytical solution given by the conditional expectation $E(f_{\sum,3}(t)|{\hat {f}}_{\sum,2}(t))$.  Therefore, given ${\hat {f}}_{\sum,3}(t)=E(f_{\sum,3}(t)|{\hat {f}}_{\sum,2}(t))$, one can express $r_2(t)$ as:

\begin{equation}
r_2(t) = {\hat {f}}_{\sum,2}(t) - {\hat {f}}_{\sum,3}(t).
\label{process5c}
\end{equation}

\noindent Generalizing this, $r_j(t)$ at scale $j$, for $j<n$, can be obtained by

\begin{equation}
r_j(t) = {\hat {f}}_{\sum,j}(t) - {\hat {f}}_{\sum,j+1}(t),
\label{estj}
\end{equation}

\noindent where
\begin{equation}
{\hat {f}}_{\sum,j+1}(t)=E(f_{\sum,j+1}(t)|{\hat {f}}_{\sum,j}(t)).
\label{condexpectation}
\end{equation}

\noindent The last residual signal is computed as $r_n(t)={\hat {f}}_{\sum,n}(t)$ to conform with the form expressed in Eq.~\ref{sumofprocesses}.  Hence, given Eq.~\ref{estj}, we have a deep cascading framework for the BRT where we can obtain the residual signal at scale $j$ (i.e., $r_j(t)$) given the previously computed ${\hat {f}}_{\sum,j}(t)$.  Furthermore, since the residual signal at scale $j-1$  (i.e., $r_{j-1}(t)$) is not involved in the computation of the residual signal at scale $j$  (i.e., $r_j(t)$) (only ${\hat {f}}_{\sum,j}(t)$ obtained from previous cascading step is), the information contained within $r_{j-1}(t)$ is not contained within $r_{j}(t)$.  As such, as scale $j$ increases, the signal information contained in ${\hat {f}}_{\sum,j}(t)$ becomes coarser and coarser, which results in residual signals $r_{j}(t)$ characterizing coarser and coarser signal information as scale increases.  Based on Eq.~\ref{estj}, the deep cascading framework for the forward Bayesian Residual Transform (BRT) is illustrated in Fig.~\ref{fig1}a.

Due to the condition of the summation of residual signals at all scales being equal to signal $f(t)$ (Eq.~\ref{sumofprocesses}), the inverse BRT is simply the summation of all residual signals $r_1(t), r_2(t), \ldots, r_n(t)$:

  \begin{equation}
  f(t) = B^{-1}(r_1(t), r_2(t), \ldots, r_n(t))=\sum_{i=1}^{n}r_i(t).
  \label{inversetransform}
  \end{equation}

\noindent The inverse Bayesian Residual Transform (inverse BRT) procedure is illustrated in Fig.~\ref{fig1}b.

\subsection{Realization of Bayesian Residual Transform via Kernel Regression}

In this study, we implement a realization of the BRT using a kernel regression strategy, which can be described as follows.  At each iteration $j$, we compute $E(f_{\sum,j+1}(t)|{\hat {f}}_{\sum,j}(t))$ (Eq.~\ref{condexpectation}) based on nonparametric Nadaraya-Watson kernel regression~\cite{Nadaraya,Watson} using a kernel function $K_j$.  Here, we employ the following Gaussian kernel function $K_j$:

     \begin{equation}
     K_j({\hat {f}}_{\sum,j}(t)-{\hat {f}}_{\sum,j}(t_i))=e^{-\frac{1}{\lambda_j^2}({\hat {f}}_{\sum,j}(t)-{\hat {f}}_{\sum,j}(t_i))^2}
     \label{kj}
      \end{equation}

\noindent Finally, the residual signal at scale $n$ (i.e., $r_n(t)$) can be set as $E(f_{\sum,n}(t)|{\hat {f}}_{\sum,n-1}(t))$, which is computed at the step where $r_{n-1}(t)$ is computed.  By setting $r_n(t)=E(f_{\sum,n}(t)|{\hat {f}}_{\sum,n-1}(t))$, the condition of the summation of signal decompositions at all scales being equal to signal $f(t)$ (i.e., Eq.~\ref{sumofprocesses}) is satisfied.  A step-by-step summary of the realization of BRT via kernel regression is shown in Algorithm~\ref{alg1}. A step-by-step summary of the inverse BRT is shown in Algorithm~\ref{alg2}.

\subsection{Noise suppression}
\label{noisesuppression}
In this study, we wish to illustrate the feasibility of utilizing the BRT for processing physiological signals through the task of noise suppression.  As such, we first establish a simple approach to noise suppression of signals using the BRT for illustrative purposes.  The noise suppression method chosen for this study is based around the idea that the observed noisy signal $f(t)$ is formed as a summation of the desired noise-free signal $f'(t)$ and an additive noise source.  Suppose that we have the true noise-free signal $f'(t)$ and we decompose it using the BRT into a series of residual signals $r_1(t), r_2(t), \ldots, r_n(t)$, where each of the residual signals characterize only information from the noise-free signal at a particular scale.  Much of the information at each scale that characterizes the noise-free signal $f'(t)$ would be concentrated within only a few of the locations in each of the residual signals.  What this means is that much of the information content related to $f'(t)$ is primarily concentrated within just a few locations at each scale.  If we were to decompose the noisy signal $f(t)$ using the BRT in a similar fashion, the locations of the residual signal at each scale that would otherwise have negligible information content associated with $f'(t)$ would now have low but not negligible information content that characterizes the noise source.  Motivated by this, we employ a noise thresholding strategy where we only keep information from locations with information content greater than the noise information content level $\theta$ at each scale.

\begin{algorithm}[h]
    \caption{Step-by-step summary for Bayesian Residual Transform via Kernel Regression}
    \begin{algorithmic}
    \REQUIRE~~\\{\STATE A signal $f(t)$}
    \STATE parameters initialization: $\lambda_1,\ldots,\lambda_{n-1}, n$
    \ENSURE~~\\{\STATE residual signals $r_1(t), r_2(t), \ldots, r_n(t)$}\\~\\
        \STATE $j=1$;
        \STATE $\hat f_{\sum,1}(t)=f(t)$;
        \WHILE {$(j < n )$}
            \STATE Compute $\hat f_{\sum,j+1}(t)=E(f_{\sum,j+1}(t)|{\hat {f}}_{\sum,j}(t))$ based on kernel regression with $K_j$ $\gets$ Eq. (\ref{kj})\
            \STATE Compute $r_j(t) = {\hat {f}}_{\sum,j}(t) - \hat f_{\sum,j+1}(t) \gets$ Eq. (\ref{estj})\
            \STATE $j=j+1$;
        \ENDWHILE
        \STATE $r_n(t) = \hat f_{\sum,j}(t)$
    \end{algorithmic}
    \label{alg1}
\end{algorithm}

Motivated by this, the noise thresholding strategy employed in this study can be described as follows.  We first perform the forward BRT on the signal $f(t)$ to obtain $n$ residual signals characterizing signal information at different scales ($r_1(t), r_2(t), \ldots, r_n(t)$).  Since the noise information content level at each scale is not known, we employ the seminal noise level estimation method proposed by Donoho~\cite{Donoho2} to determine the noise threshold $\theta$ at each scale, which can be described as follows.  At scale $j$, we estimate the noise threshold $\theta_j$ at each scale $j$ using the noise-adaptive scale estimate, which can be expressed by:

\begin{equation}
\theta_j = MAD(r_j) / {\Phi^{-1}(3/4)},
\label{noiseest2}
\end{equation}

\noindent where $MAD$ is the median absolute deviation and $\Phi^{-1}$ is the normal inverse cumulative distribution function.  Based on $\theta_j$, noise thresholding is achieved to obtain noise-suppressed residual signal $r'_j(t)$ by:

\begin{align}
\hspace*{0.25in} r'_j(t)&=\left\{\begin{array}{ccccc}
0       & \textup{if } & |r_j(t)| < & \theta_j
\\
r_j(t) & \textup{if } &                  & otherwise &
\\
\end{array}
\right.
\label{eqn-2}
\end{align}

\begin{figure*}[tp]
	\centering
    \includegraphics[width=0.7\linewidth]{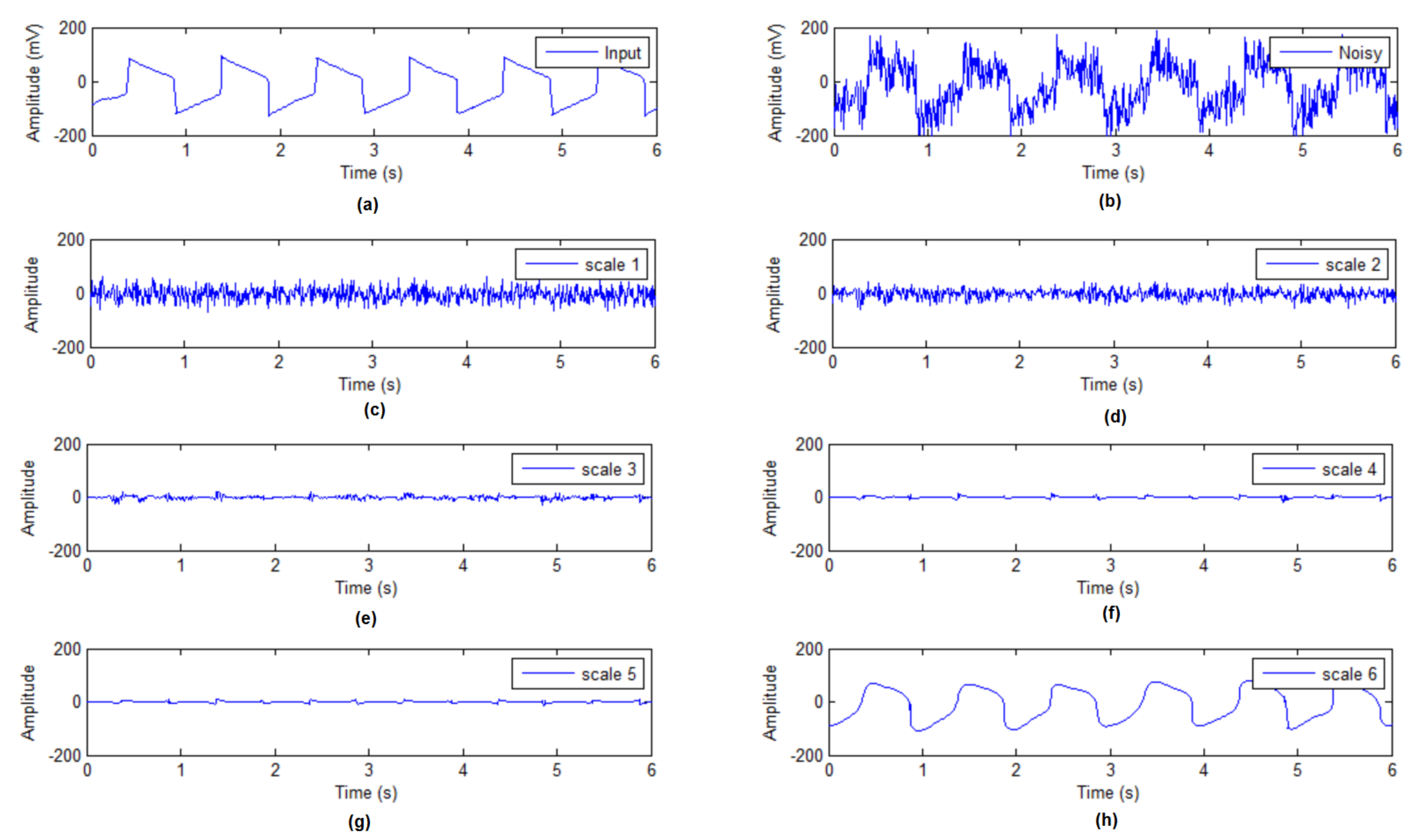}
	\caption{Example of multi-scale signal decomposition using the BRT. \textbf{(a)} Baseline periodic test signal. \textbf{(b)} Noisy input signal with zero-mean Gaussian noise. \textbf{(c)-(h)} Signal decompositions using the BRT at different scales. It can be observed that the noise process contaminating the test signal is well characterized in the decompositions at the lower (finer) scales (scales 1 to 3), while the structural characteristics of the test signal is well characterized in the decompositions at the higher (coarser) scales (scales 4 to 6). }
	\label{fig2}
\end{figure*}

\noindent Finally, the inverse BRT (Eq.~\ref{inversetransform}) is performed on the set of $n$ noise-suppressed residual signals at the different scales ($r'_1(t), r'_2(t), \ldots, r'_n(t)$) to produce the noise-suppressed signal $f'(t)$.  A step-by-step summary of the noise suppression method using the BRT is shown in Algorithm~\ref{alg3}.\\
\begin{algorithm}[h]
    \caption{Step-by-step summary for inverse Bayesian Residual Transform}
    \begin{algorithmic}
    \REQUIRE~~\\{\STATE residual signals $r_1(t), r_2(t), \ldots, r_n(t)$}
    \STATE parameters initialization: $n$
    \ENSURE~~\\{\STATE A signal $f(t)$}\\~\\
        \STATE $j=1$;
        \STATE $f(t)=0$;
        \WHILE {$(j \leq n )$}
            \STATE $f(t)=f(t)+r_j(t)$;
            \STATE $j=j+1$;
        \ENDWHILE
    \end{algorithmic}
    \label{alg2}
\end{algorithm}
\begin{algorithm}[h]
    \caption{Step-by-step summary for noise suppression using Bayesian Residual Transform}
    \begin{algorithmic}
    \REQUIRE~~\\{\STATE A noisy signal $f(t)$}
    \STATE parameters initialization: $n$
    \ENSURE~~\\{\STATE A noise-suppressed signal $f'(t)$}\\~\\
        \STATE Perform the BRT on $f(t)$ to obtain residual signals $r_1(t), r_2(t), \ldots, r_n(t)$. $\gets$ Algorithm~\ref{alg1}
        \STATE $j=1$;
        \WHILE {$(j \leq n )$}
            \STATE Compute noise threshold $\theta_j$ $\gets$ Eq.~\ref{noiseest2};
            \STATE Compute noise-suppressed residual signal $r'_j(t)$ via threshold using $\theta_j$ $\gets$ Eq.~\ref{eqn-2};
            \STATE $j=j+1$;
        \ENDWHILE
        \STATE Perform inverse BRT on $r'_1(t), r'_2(t), \ldots, r'_n(t)$ to obtain noise-suppressed signal $f'(t)$ $\gets$ Algorithm~\ref{alg2}
    \end{algorithmic}
    \label{alg3}
\end{algorithm}

\begin{figure*}[tp]
	\centering
    \includegraphics[width=0.8\linewidth]{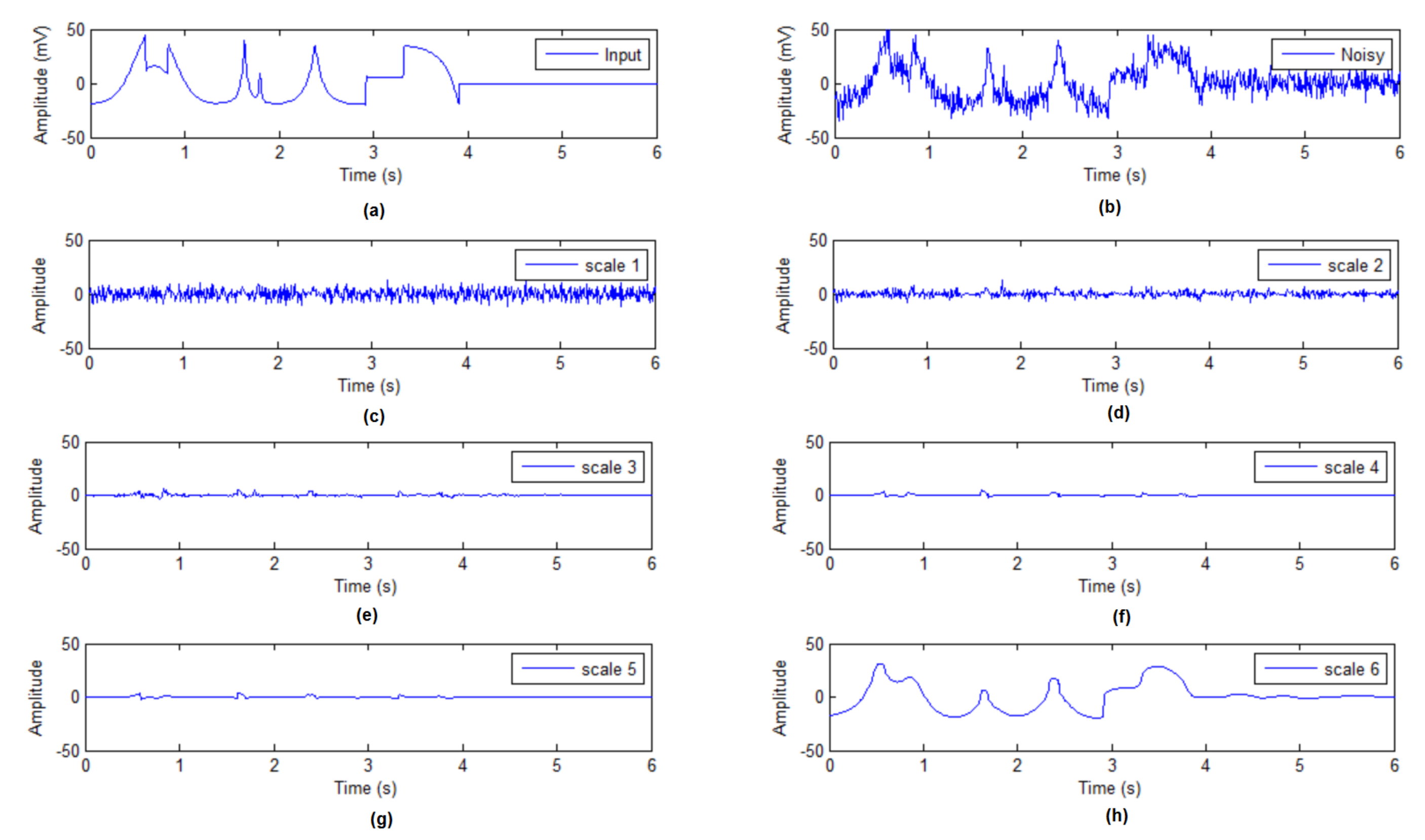}
	\caption{Example of multi-scale signal decomposition using the BRT. \textbf{(a)} Baseline piece-wise regular test signal. \textbf{(b)} Noisy input signal with zero-mean Gaussian noise. \textbf{(c)-(h)} Signal decompositions using the BRT at different scales. It can be observed that, as with the periodic signal example, the noise process contaminating the signal is well characterized in the decompositions at the lower (finer) scales (scales 1 to 2), while the structural characteristics of the test signal is well characterized in the decompositions at the higher (coarser) scales (scales 3 to 6). Furthermore, more noticeable here than in the periodic example, it can be seen that that the decomposition at each scale exhibits good signal structural localization.}
	\label{fig3}
\end{figure*}

\section{Experimental setup}
\label{setup}

In this study, to illustrate the feasibility of utilizing the BRT for processing physiological signals, we performed a SNR analysis using electrocardiography (ECG) signals to study the performance of the BRT for the task of noise suppression.  ECG signals from the MIT-BIH Normal Sinus Rhythm Database~\cite{Sinus} were used in this study to perform the SNR analysis.  This database consists of 18 ECG recordings (recorded at a sampling rate of 128 Hz) of subjects conducted at the Arrhythmia Laboratory in the Beth Israel Deaconess Medical Center.  The subjects were found to have no significant arrhythmias.  A total of 18 low-noise segments of 10 seconds was extracted, one from each recording, based on visual inspection to act as the baseline signals for evaluation.  To study noise suppression performance at different SNR levels, each of the 18 baseline signals were contaminated by white Gaussian noise to produce noisy signals with SNR ranging from 12 dB to 2.5 dB (with 20 different noisy signals at each SNR), resulting in 3960 different signal perturbations used in the analysis.  For comparison purposes, wavelet denoising methods with the following shrinkage rules were also used: i) Stein's Unbiased Risk (SURE)~\cite{Donoho3}, ii) Heuristic SURE (HSURE)~\cite{matlab}, iii) Universal (UNI)~\cite{matlab}, and iv) Minimax (MINIMAX)~\cite{Donoho2}.  Each of the methods uses their corresponding noise threshold and shrinkage rules in the original works.  To quantitative evaluate noise suppression performance, we compute the SNR improvement as follows~\cite{Akhbari}:

\begin{equation}
SNRI = 10\log\left(\frac{\sum_t(f(t) - f_b(t))^2}{\sum_t(f'(t) - f_b(t))^2}\right)
\end{equation}
\noindent where $f(t)$, $f_b(t)$, and $f'(t)$ are the noisy, baseline, and noise-suppressed signals obtained using a noise suppression method, respectively.\\
~\\
To study the effect of the number of scales $n$ on noise suppression performance, the same SNR analysis is performed as described above for $n=\{2,3,4,5,6\}$.\\

\subsection{Implementation details}

The BRT is implemented in MATLAB (The MathWorks, Inc.), with the nonparametric conditional expectation estimates implemented in C++ and compiled as a dynamically linked MATLAB Executable (MEX) to improve computational speed.  The only free parameters of the implemented realization of the BRT are the standard deviations used to model the residual signals (e.g., $\lambda$), the number of scales $n$, and time window size, which can be adjusted by the user to find a tradeoff between noise suppression quality and computational costs.  For the SNR analysis of ECG signals, $\lambda$ is set equally for all scales to the standard deviation of $f(t)$ for simplicity, $n$ is set at 6 scales, and the time window size is set to $0.1s$.  For this configuration, the current implemented realization of the BRT can process a 1028-sample signal in $<$1 second on an Intel(R) Core(TM) i5-3317U CPU at 1.70GHz CPU.  For the wavelet-based methods tested (SURE, HSURE, UNI, and MINIMAX), as implemented in MATLAB (The MathWorks, Inc.), soft thresholding with the Coiflet3 mother wavelet at 6 scales and single level rescaling was used as it was found to provide superior results for ECG noise suppression~\cite{Sameni}.  Each of the methods uses their corresponding noise threshold and shrinkage rules as specified in the original works.

\section{Experimental Results}
\label{results}

To illustrate the feasibility of utilizing the BRT for processing physiological signals, such as for the task of noise suppression, we first performed the BRT on two test signals: i) a noisy periodic test signal, and ii) a noisy piece-wise regular test signal.  The multi-scale signal decomposition using the BRT on a noisy periodic test signal is shown in Fig.~\ref{fig2}.  Here, a baseline test signal (Fig.~\ref{fig2}a) is contaminated by a zero-mean Gaussian noise process to produce a noisy signal (Fig.~\ref{fig2}b) and then decomposed using the BRT at different scales (Figs.~\ref{fig2}c-h).  It can be observed that the noise process contaminating the signal is well characterized in the decompositions at the lower (finer) scales (scales 1 to 3), while the structural characteristics of the test signal is well characterized in the decompositions at the higher (coarser) scales (scales 4 to 6).

\begin{figure*}[tp]
	\centering
    \includegraphics[width=0.9\linewidth]{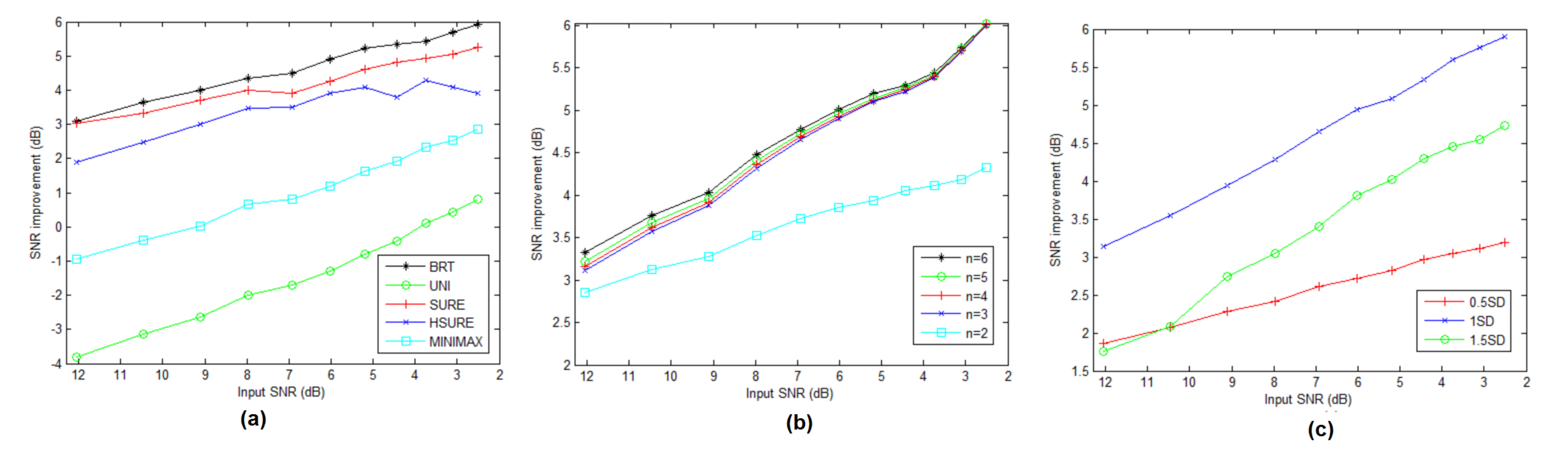}
	\caption{Application of the BRT on ECG signals. \textbf{(a)} A plot of the mean SNR improvement vs. the different input SNRs ranging from 12 dB to 2.5 dB for the MIT-BIH Normal Sinus Rhythm Database for the tested methods.  Noise-suppression method using the BRT provided strong SNR improvements across all SNRs, with performance comparable to SURE and higher than the other 3 tested methods. \textbf{(b)} A plot of the mean SNR improvement vs.  the different input SNRs ranging from 12 dB to 2.5 dB for the method using the BRT with different number of scales $n$. \textbf{(c)} A plot of the mean SNR improvement vs. the different input SNRs ranging from 12 dB to 2.5 dB for the method using the BRT with different multiples of the standard deviation (SD) for $\lambda$.}
	\label{fig4}
\end{figure*}

The multi-scale signal decomposition using the BRT on a noisy piece-wise regular test signal (generated using~\cite{tp}) is shown in Fig.~\ref{fig3}.  As with the previous example, a baseline test signal (Fig.~\ref{fig3}a) is contaminated by a zero-mean Gaussian noise process to produce a noisy signal (Fig.~\ref{fig3}b) and then decomposed using the BRT at different scales (Figs.~\ref{fig3}c-h).  It can be observed that, as with the periodic signal example, the noise process contaminating the signal is well characterized in the decompositions at the lower (finer) scales (scales 1 to 2), while the structural characteristics of the test signal is well characterized in the decompositions at the higher (coarser) scales (scales 3 to 6). Furthermore, more noticeable here than in the periodic signal example, it can be seen that that the decomposition at each scale exhibits good signal structural localization.  Therefore, given the ability of the BRT to decouple the noise process from the true signal into different scales, as illustrated in both the periodic and piece-wise regular test signals, the BRT has the potential to be useful for performing noise suppression on signals while preserving inherent signal characteristics.

In this study, to illustrate the feasibility of utilizing the BRT for processing physiological signals, we introduced a simple thresholding approach to noise suppression using the BRT for illustrative purposes (\textbf{see Section~\ref{noisesuppression}}).  We then performed a quantitative SNR analysis using electrocardiography (ECG) signals from the MIT-BIH Normal Sinus Rhythm Database~\cite{Sinus} to study the performance of the BRT for the task of noise suppression, where the SNR improvement (\textbf{see Section~\ref{noisesuppression} for formulation}).

A plot of the mean SNR improvement of the tested methods vs. the different input SNRs ranging from 12 dB to 2.5 dB is shown in Fig.~\ref{fig4}a.  It can be observed that the noise-suppression method using the BRT provided strong SNR improvements across all SNRs, comparable to SURE and higher than the other 3 tested methods.  It can also be observed that the UNI method consistently achieved SNR improvements below 0 dB.  This is primarily due to the tendency to overestimate the noise level, resulting in signal oversmoothing and thus producing a noise-suppressed signal that is less similar to the baseline signal than the actual noisy signal.  It can also be observed that the SNR improvement increases as the SNR of the input noisy signal decreases, which indicates that greater benefits are obtained through the use of noise suppression methods in low signal SNR scenarios.

To study the effect of the number of scales $n$ on noise suppression performance, a plot of the mean SNR improvement vs. the different input SNRs ranging from 12 dB to 2.5 dB for the method using the BRT with a range of different number of scales $n$ is shown in Fig.~\ref{fig4}b.  It can be observed that a significant gain in SNR improvement exists going from $n=2$ to $n=3$, with smaller SNR improvement gains from $n=3$ all the way to $n=6$.  Furthermore, it can be observed that the SNR improvement gains from increasing the number of scales become smaller and smaller as the input SNR decreases, with the SNR improvement for $n=3$ to $n=6$ being approximately the same when the input SNR is 2.5 dB.  Therefore, this indicates that the effect of selecting the number of scales on noise suppression performance can be significant and thus a balance between SNR improvement and the computational complexity of the BRT (which grows linearly with the number of scales) is necessary, particularly given the SNR of the noisy signal.

To study the effect of the standard deviation (SD) used for $\lambda$ on noise suppression performance, a plot of the mean SNR improvement vs. the different input SNRs ranging from 12 dB to 2.5 dB for the method using the BRT with a range of different multiples of SD used for $\lambda$ is shown in Fig.~\ref{fig4}c.  It can be observed that a significant gain in SNR improvement exists going from $0.5SD$ to $1SD$, with a significant drop in SNR improvements going from $1SD$ to $2SD$.  Furthermore, it can be observed that there are noticeable SNR improvement gains going from $0.5SD$ to $2SD$ that grows larger as the input SNR decreases.  Therefore, this indicates that the effect of selecting $\lambda$ on noise suppression performance can be significant, and careful selection may be important when dealing with different types of signals.  For the signals tested here, it was found that $1SD$ provided the strongest results.

\begin{figure*}[tp]
	\centering
    \includegraphics[width=0.8\linewidth]{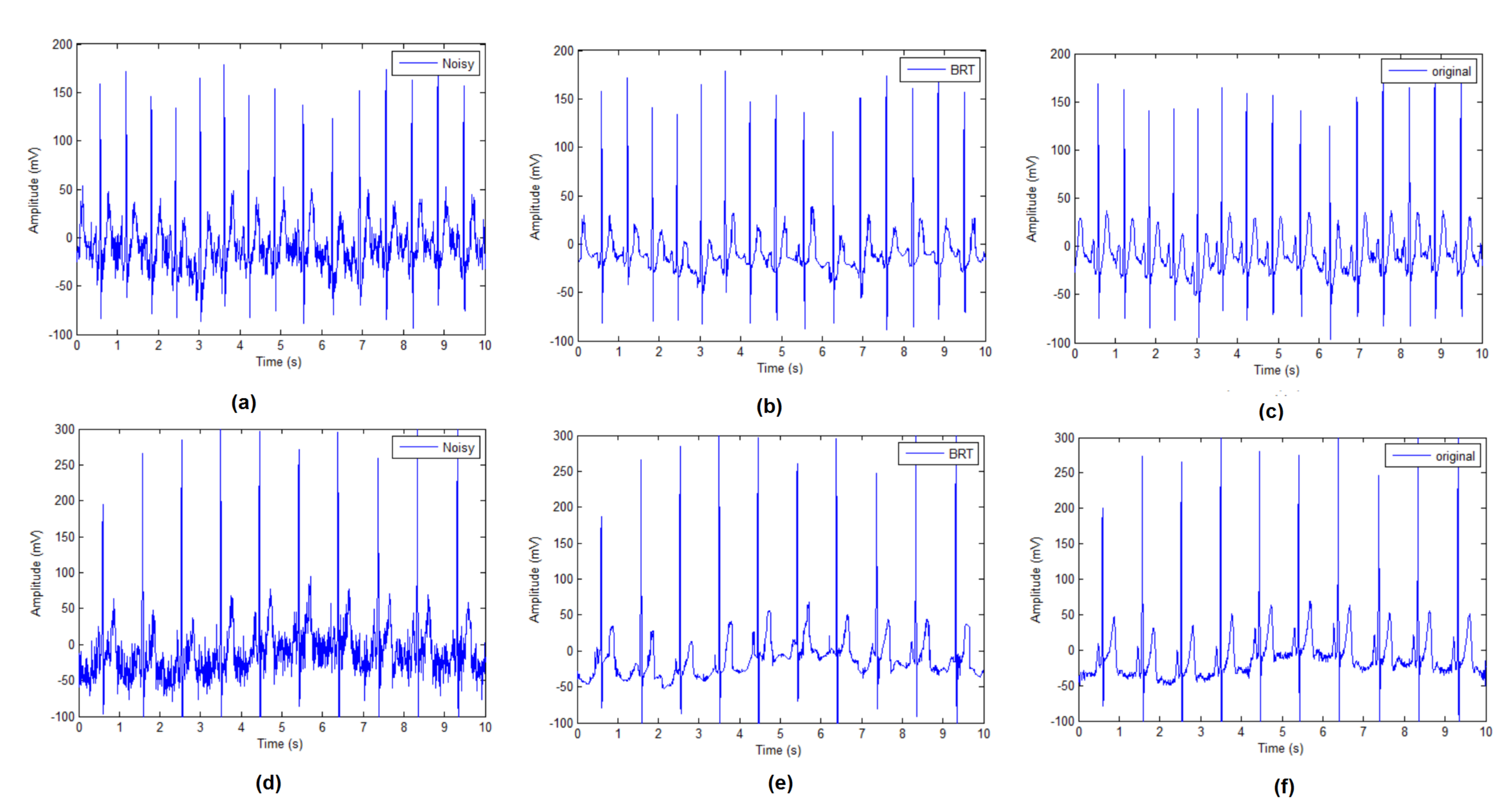}
	\caption{Application of the BRT on ECG signals. \textbf{(a)} Noisy input signal with SNR=12 dB, \textbf{(b)} noise-suppressed results using BRT for \textbf{a}, and \textbf{(c)} the corresponding original signal. \textbf{(d)} Another noisy input signal with SNR=12 dB, and \textbf{(e)} noise-suppressed results using BRT for \textbf{d}, and \textbf{(f)} the corresponding original signal. The results produced using the BRT has significantly reduced noise artifacts while the signal characteristics are preserved. }
	\label{fig5}
\end{figure*}

Typical results of noise-suppressed signals produced by the method using the BRT are shown in Fig.~\ref{fig5}b and Fig.~\ref{fig5}e (corresponding to two different 12 dB noisy input signals shown in Fig.~\ref{fig5}a and Fig.~\ref{fig5}d, respectively).  Visually, it can be seen that the BRT was effectively used to produce signals with significantly reduced noise artifacts while preserving signal characteristics.  Results in this study show that it is feasible to utilize the BRT for processing physiological signals for tasks such as noise suppression.

\section{Conclusion}
\label{conclusions}

In this study, the feasibility of employing a Bayesian-based approach to multi-scale signal decomposition introduced here as the Bayesian Residual Transform for use in the processing of physiological signals.  The Bayesian Residual Transform decomposes a signal into a set of residual signals, each characterizing information from the signal at different scales and following a particular probability distribution.  This allows information at different scales to be decoupled for the purpose of signal analysis and, for the purpose of noise suppression, allows for information pertaining to the noise process contaminating the signal to be separated from the rest of the signal characteristics.  This trait is important for performing noise suppression on signals while preserving inherent signal characteristics.  SNR analysis using a set of ECG signals from the MIT-BIH Normal Sinus Rhythm Database at different noise levels demonstrated that it is feasible to utilize the BRT for processing physiological signals for tasks such as noise suppression.

Given the promising results, we aim in the future to investigate alternative adaptive thresholding schemes for the task of noise suppression in physiological signals characterized by nonstationary noise, so that one can better adapt to the nonstationary noise statistics embedded at different scales.  Moving beyond low-level signal processing tasks such as noise suppression, we aim with our future work to investigate and devise methods for multi-scale analysis of a signal using the Bayesian Residual Transform, which could in turn lead to improved features for signal classification.  Finally, we aim to investigate the extension and generalization of the Bayesian Residual Transform for dealing with high-dimensional physiological signals such as vectorcardiographs (VCG)~\cite{Sameni2}, and dealing with high-dimensional medical imaging signals from systems such as multiplexed optical high-coherence interferometry~\cite{Farnoud}, optical coherence tomography~\cite{OCT1,OCT2}, dermatological imaging~\cite{derm}, diffusion weighted magnetic resonance imaging (DWI)~\cite{Koh,Bihan,Shafiee}, microscopy~\cite{microscopy1,microscopy2}, dynamic contrast enhanced MRI (DCE-MRI), and correlated diffusion imaging~\cite{Wong,dualstage}.

\section{Acknowledgment}

This work was supported by the Natural Sciences and Engineering Research Council of Canada, Canada Research Chairs Program, and the Ontario Ministry of Research and Innovation.

\ifCLASSOPTIONcaptionsoff
  \newpage
\fi




\begin{IEEEbiography}[{\includegraphics[width=1in,height=1.25in,clip,keepaspectratio]{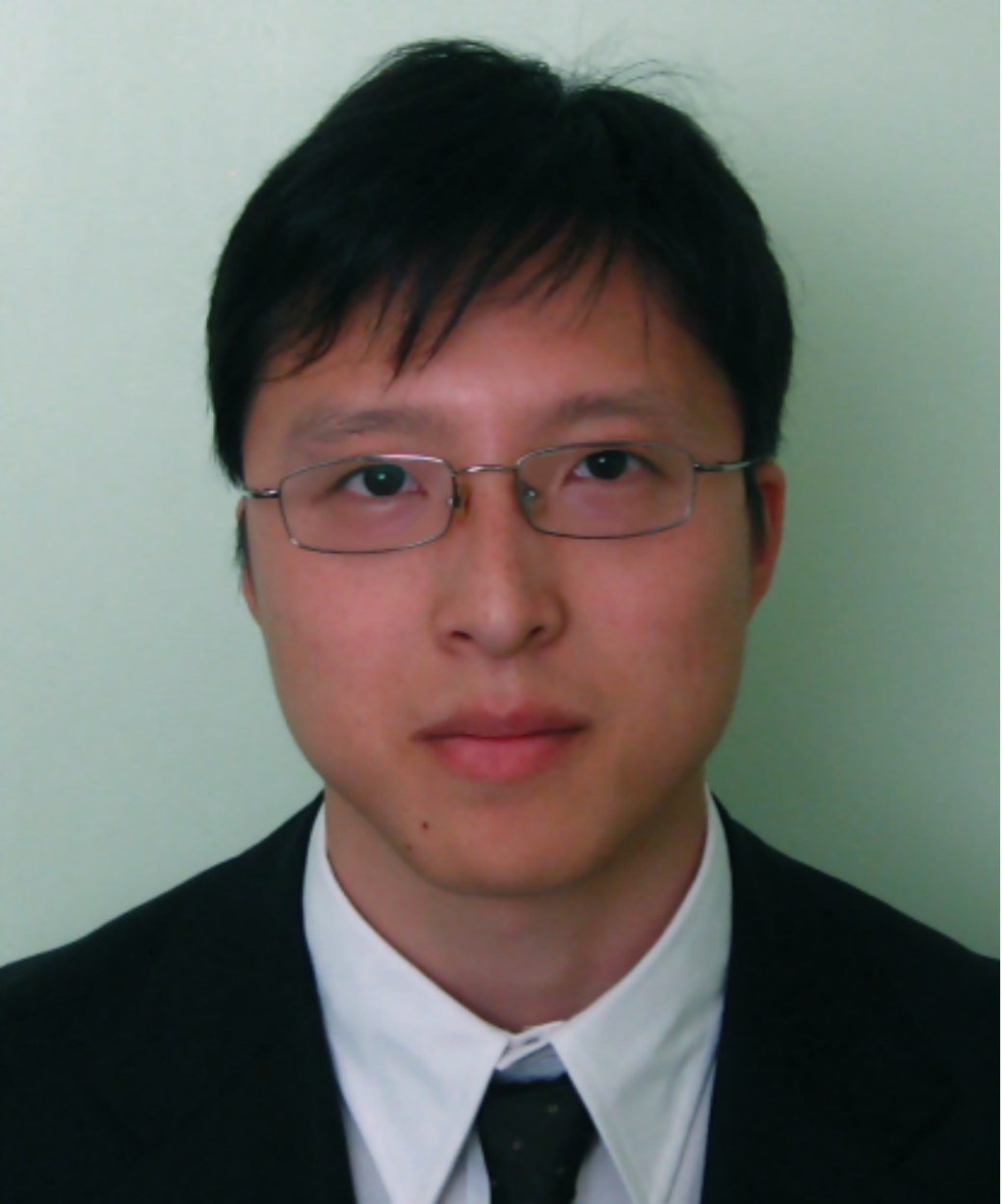}}]{Alexander Wong} (M' 05) received the B.A.Sc. degree in Computer Engineering from the University of Waterloo, Waterloo, ON, Canada, in 2005, the M.A.Sc. degree in Electrical and Computer Engineering from the University of Waterloo, Waterloo, ON, Canada, in 2007, and the Ph.D. degree in Systems Design Engineering from the University of Waterloo, ON, Canada, in 2010. He is currently the Canada Research Chair in Medical Imaging Systems, Co-director of the Vision and
Image Processing Research Group, and an Assistant Professor in the Department of Systems Design Engineering, University of Waterloo, Waterloo, Canada. He has published refereed journal and conference papers, as well as patents, in various fields such as computer vision, graphics, image processing, multimedia systems, and wireless communications. His research interests revolve around imaging, image processing, computer vision, pattern recognition, and cognitive radio networks, with a focus on integrative biomedical imaging systems design, probabilistic graphical models, biomedical and remote sensing image processing and analysis such as image registration, image denoising and reconstruction, image super-resolution, image segmentation, tracking, and image and video coding and transmission.  Dr. Wong has received two Outstanding Performance Awards, an Engineering Research Excellence Award, an Early Researcher Award from the Ministry of Economic Development and Innovation, two Best Paper Awards by the Canadian Image Processing and Pattern Recognition Society (CIPPRS), a Distinguished Paper Award from Society for Information Display, and the Alumni Gold Medal.
\end{IEEEbiography}

\begin{IEEEbiography}[{\includegraphics[width=1in,height=1.25in,clip,keepaspectratio]{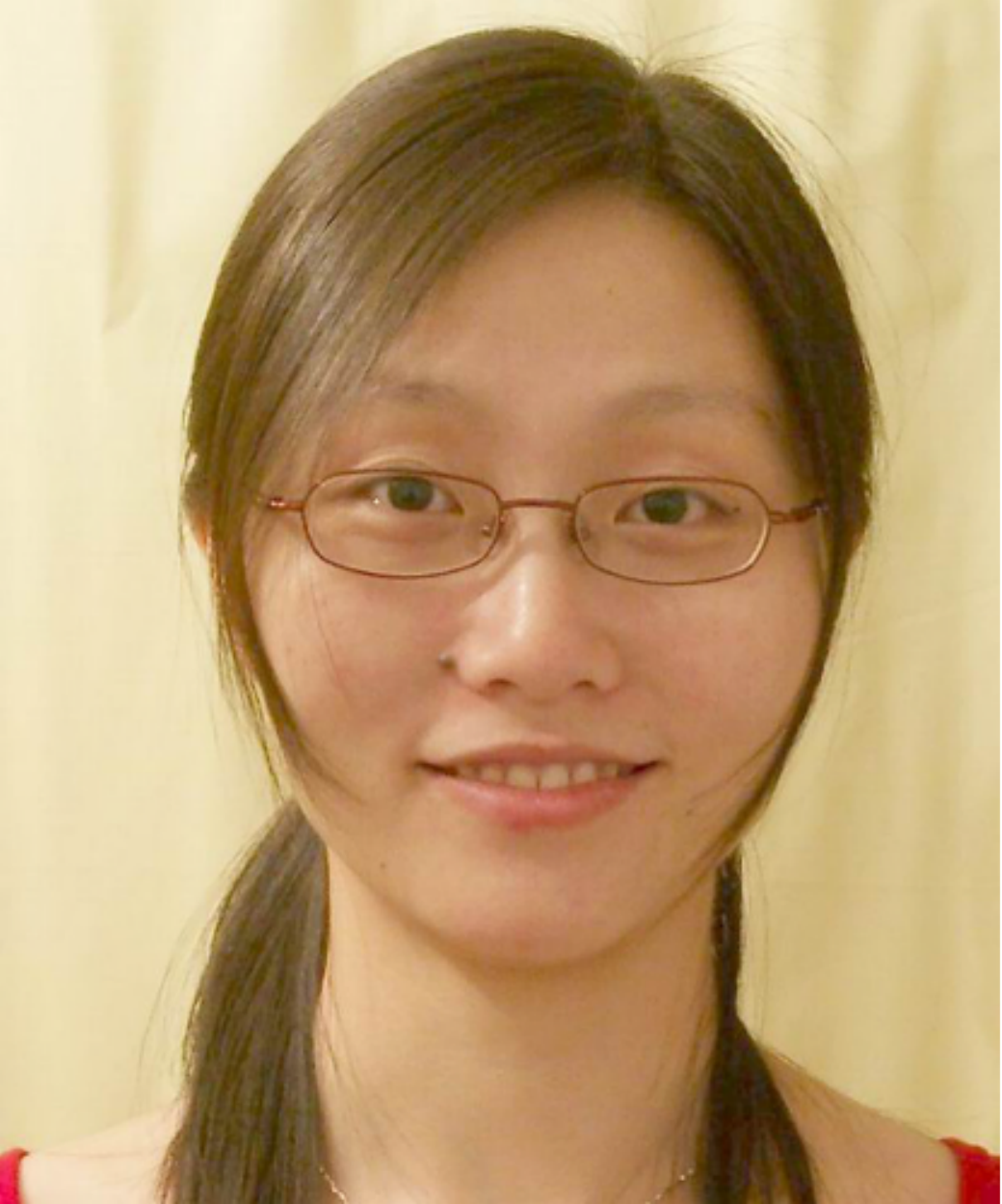}}]{Xiao Yu Wang} received the M.A.Sc. degree in
Electrical Engineering from Concordia University,
Montreal, Canada, in 2006, and the Ph.D. degree in Electrical and Computer Engineering from the University of Waterloo, ON, Canada, in 2011. She is currently an Adjunct Assistant Professor in the Department of Systems Design Engineering, University of Waterloo,
Waterloo, Canada.  Her research interests include stochastic graphical learning and modeling for large-scale networks and data mining and visualization, affective computing, image processing, computer vision, signal processing, femtocell networking, network control theory, wideband spectrum sensing, and dynamic spectrum access.  Her current focus is on efficient high-resolution, remote spatial biosignals measurements using video imaging for affective computing.
\end{IEEEbiography}

\end{document}